\documentclass{webofc}
\usepackage[varg]{txfonts}   
\begin{document}
\title{Demonstrating the Principles of Aperture Synthesis with Table-Top Laboratory Exercises}
%
%

\author{\firstname{J.\ M.} \lastname{Marr}\inst{1}\fnsep\thanks{\email{marrj@union.edu}} \and
        \firstname{A.\ E.\ E.} \lastname{Rogers}\inst{2} \and
        \firstname{V.\ L.} \lastname{Fish}\inst{2} \and
        \firstname{F.\ P.} \lastname{Wilkin}\inst{1} \and
				\firstname{M.\ B.} \lastname{Arndt}\inst{3} \and
				\firstname{G.} \lastname{Holodak}\inst{1} \and
        \firstname{K.} \lastname{Durkota}\inst{1}
				}

\institute{Union College, Schenectady, NY, USA 
\and
           MIT Haystack Observatory, Westford, MA, USA 
\and
           Bridgewater State University, Bridgewater, MA, USA
          }

\abstract{
Many undergraduate radio astronomy courses are unable to give a detailed treatment of aperture synthesis due to time constraints and limited math backgrounds of students.  We have taken a laboratory-based approach to teaching radio interferometry using a set of college-level, table-top exercises. These are performed with the Very Small Radio Telescope (VSRT), an interferometer developed at the Haystack Observatory using satellite TV electronics as detectors and compact fluorescent light bulbs as microwave signal sources. The hands-on experience provided by the VSRT in these labs allows students to gain a conceptual understanding of radio interferometry and aperture synthesis without the rigorous mathematical background traditionally required. 

The data are quickly and easily processed using a user-friendly data analysis Java package, VSRTI\_Plotter.jar.  This software can also be used in the absence of the equipment as an interactive computer activity to demonstrate an interferometer's responses to assorted surface brightness distributions.  The students also gain some familiarity with Fourier transforms and an appreciation for the Fourier relations in interferometry using another Java package, the Tool for Interactive Fourier Transforms (TIFT).  We have successfully used these tools in multiple offerings of our radio astronomy course at Union College
}

\maketitle

\section{Introduction}
\label{intro}

The radio astronomical technique of aperture synthesis, in which high-resolution images are produced from interferometer arrays, has been a productive tool for astronomers for decades. From its initial development in the 1950s and early 1960s, aperture synthesis was quickly recognized as a significant advancement for science.  Sir Martin Ryle shared the Nobel Prize for Physics in 1974 ``for his observations and inventions, in particular of the aperture synthesis technique''\,\cite{Nobel}. The best resolution attained with single-dish radio telescopes has historically been of order 20 arcseconds, while that achieved with aperture synthesis can be four orders of magnitude better.  With the subsequent construction of the (now renamed) Jansky Very Large Array\footnote{The Very Large Array is operated by the National Radio Astronomy Observatory, which is a facility of the National Science Foundation operated under cooperative agreement by Associated Universities, Inc.} and continent-scale arrays for very long baseline interferometry, radio astronomers have been able to produce images with much higher angular resolution than has been possible at other wavelengths. 
Yet, because of the complexity of the math involved, aperture synthesis is often excluded from undergraduate curricula in physics and astronomy.

We introduce here a set of labs at the undergraduate level which provide hands-on experience with the basics of aperture synthesis observations and how the data reveal information about the size and structure of the observed sources. These labs are designed to give students a conceptual understanding of aperture synthesis without the need for the dense mathematical formalism that usually accompanies the topic in graduate-level classes.

\section{The basic concept of aperture synthesis} \label{sec:basicconcept}

Successful completion of the labs does not require that the students, or instructors, know the mathematical formalism.
As such, in this section, we give a brief discussion of what an aperture synthesis observation entails, touching on only those aspects necessary to comprehend the principles demonstrated in these labs. 
Details about interferometry can be found in textbooks \,\cite{MSK,WRH,BGS,TMS,Kraus}.

\begin{figure}[h]
\centering
\scalebox{0.7}{\includegraphics{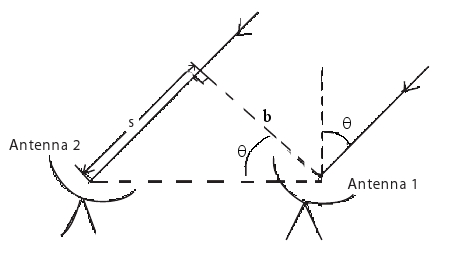}}
\caption{A pair of antennas receive radiation from a celestial source in a direction $\theta$ relative to the mid-plane position. The path from the source to antenna 2 contains the extra distance $s$, which causes a phase difference when the signals are combined.  The projected baseline (component perpendicular to the source direction) is shown as the length $b$. \label{fig:pathdiff}}
\end{figure}

In aperture synthesis a number of antennas, arranged in a particular pattern, or ``array,'' receive the radiation from a celestial source simultaneously and the signals are combined pairwise.  The method of signal combination in most modern interferometers is a cross-correlation, but the signals can also be added.  The response of each pair of antennas contains an amplitude and a phase which, customarily, are represented as a complex number.  (The ``amplitude'' in a radio interferometer's output is called the ``modulus'' of a complex number in standard mathematical usage.)  For a single point of emission, the amplitude is proportional to the flux density of the source while the phase is related to the difference in path lengths to the two antennas, as depicted in Figure~\ref{fig:pathdiff}.  For a general source of arbitrary structure, the detected amplitude and phase are the complex superposition of the responses due to each unresolved point of emission within the field of view.

In an aperture synthesis observation, the fully calibrated response of each pair of antennas, termed a ``visibility,'' is a function of the separation of the antennas, generally referred to as the ``baseline,'' $\vec{b}$. If the source is not located along the mid-plane of the baseline, $\vec{b}$ is the component of the baseline perpendicular to the direction of the source.

The visibility for any particular antenna pair is most sensitive to source structure on an angular scale proportional to $\lambda/b$, where $\lambda$ is the wavelength of the observation. This is a familiar concept from optics, in which the resolution of a telescope is proportional to $\lambda/D$, where $D$ is the diameter of the objective lens or mirror.  With an array of more than two antennas, the visibility for each pair is obtained and hence the visibility function can be probed over a wide range of baseline lengths and orientations.  In this way, information about the source structure on multiple angular scales enables one to produce an image of the source.  When written as complex values, the visibility function is related to the image of the source via a two-dimensional Fourier transform.  
 
In the labs described here, the presentation of the important principles is simplified in two ways. First, since the Very Small Radio Telescope (VSRT) measures only the amplitude of the visibilities and not the phase, complex numbers can be avoided.  Second, the labs are performed with all sources and antennas in the horizontal plane and so the analysis is reduced to one-dimension.  Students are introduced to the visibility function by measuring the interferometer response as a function of antenna separation and discover the relationship between visibilities and simple source structure.  By extrapolation, the students gain an appreciation of how information about source structure can be recovered from data obtained with many pairs of antennas.

The visibility as a function of baseline is analogous to the diffraction and interference patterns produced in the single and double slit experiments.  Radiation from each point in the observed source is detected by each antenna and the different path lengths result in constructive or destructive interference depending on the point's position in the sky and the baseline geometry when the signals are combined.  For a source of arbitrary brightness pattern, the resulting visibility function is the sum of the interferometer responses for all points of emission in the source.  When written as complex values, the visibility function, with $\vec{b}/\lambda$ as the independent variable, is related to the image of the source, i.e. intensity as a function of angle, $\vec{\theta}$, on the sky, via a Fourier transform.  In practice, the radio astronomer obtains an image of a source with software which performs an inverse Fourier transform on the visibility data.   

\section{Equipment}
\label{sec:VSRT}

\begin{figure}[h]
\centering
\scalebox{0.1}{\includegraphics{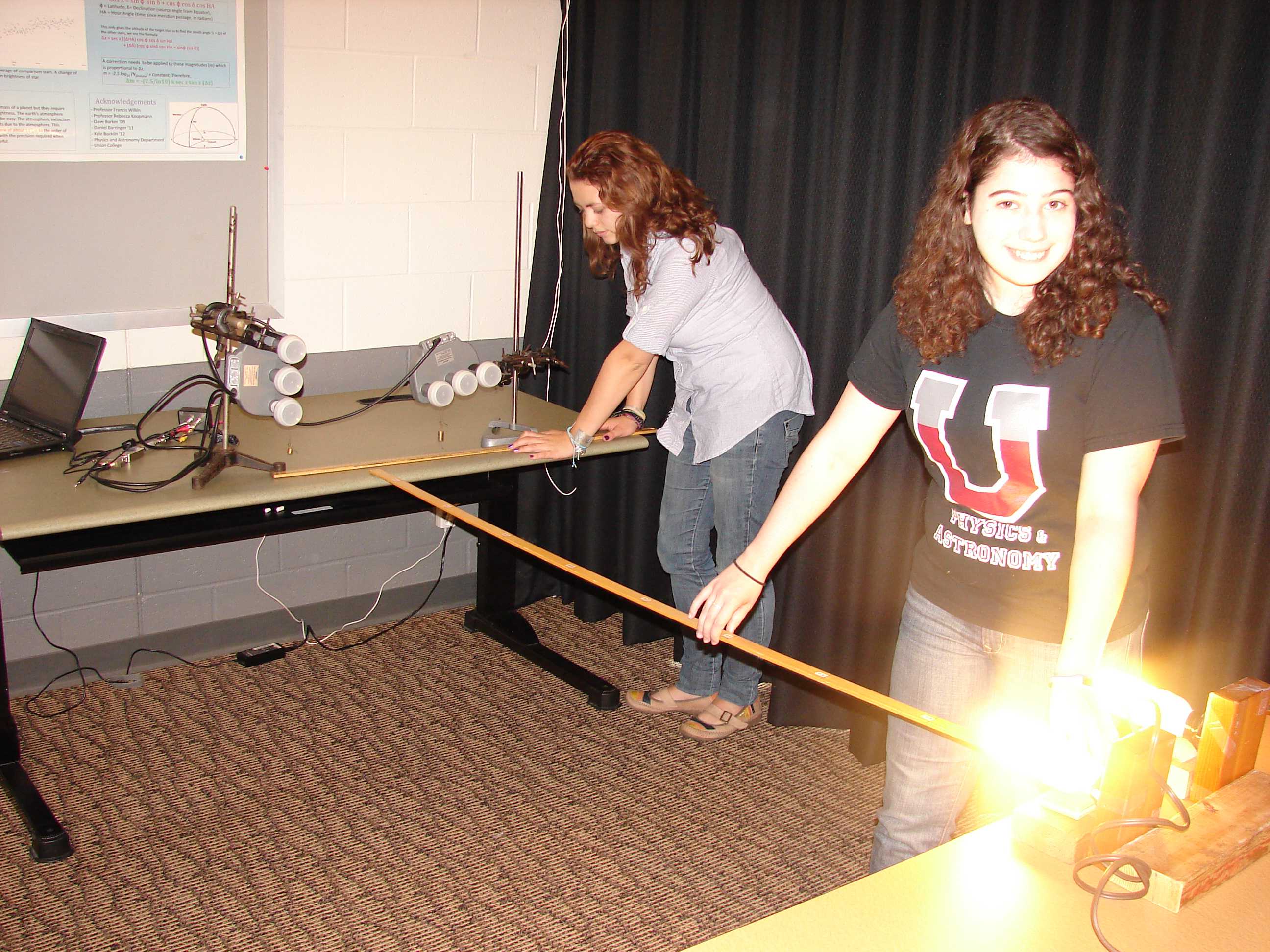}}
\caption{\label{fig:setup} Union College students working with the VSRT.  Two ``triple'' DirecTV feeds, taken from TV satellite dishes, used here as radio antennas, receive and detect the radio emission from compact fluorescent light bulbs (CFLs). Only one feed of each triple is active.  The CFLs can be moved to assorted separation distances.  The output spectrum from the interferometer is displayed on the laptop screen, and the data files are recorded.}
\end{figure}

The laboratory exercises described herein use the VSRT (Very Small Radio Telescope), a laboratory interferometer developed by MIT Haystack Observatory.  A description of the operation of the VSRT as an additive interferometer is provided online \,\cite{AR}.
Composed of commercially-available electronics and satellite TV equipment, the VSRT is a low-cost instrument designed to demonstrate the principles of radio waves and interferometry in high school and college-level labs \,\cite{DFN}.  The VSRT is stable, reliable, and easy to assemble, operate and manipulate in the lab room.  (Full details of the system can be found at the VSRT project web page\,\cite{VSRTurl}.)
The VSRT uses Ku-band satellite TV feeds to receive radiation near 12 GHz. The feeds can be installed in satellite dishes to increase the collecting area and directionality of the instrument, as is done for an experiment measuring the solar diameter \,\cite{Fish}. Alternatively, the feeds can be used without the dishes to observe strong radio sources in the lab. The exercises discussed here use the instrument in this manner, with compact fluorescent light bulbs (CFLs) serving as radio sources (see Figure \ref{fig:setup}).  

In addition to visible light, CFLs produce broadband radio emission from 100 MHz up to 100 GHz when the free electrons in the plasma collide with the glass walls of the bulb \,\cite{MS}. The CFLs can be hidden by an optically-opaque material that is transparent to radio waves (e.g., a cardboard box) to demonstrate that the feeds are sensitive to the bulbs' radio emission, not their visible light.

\section{Laboratory Exercises}
\label{sec:labs}

We now describe the labs, designed to help students develop an intuitive sense of how an array of antennas can be used to infer aspects of the spatial structure of radio sources. To facilitate data analysis, we have produced a package of Java programs, named ``VSRTI\_Plotter,'' available online\,\cite{plotter,VSRTurl}.  These programs also have links to lab instructions and can produce overlays of theoretical models with adjustable parameters.  
For the sake of simplicity the lab set-ups occur entirely in the horizontal plane so that the maps of the sources, as seen from the position of the feeds, are simple plots of intensity vs.\ position in the horizontal direction.  

\subsection{The primary beam}
\label{sec:beam}

Astronomy students are usually familiar with the fact that the angular resolution of a single telescope, ignoring the atmosphere, is limited by diffraction and is approximately $\lambda/D$.  With interferometers, the central diffraction peak of each individual telescope is known as the ``primary beam.''  When an astronomical source is not at the center of the primary beam, the detected power is decreased.  The primary beam size of the individual antennas of an array, therefore, places an effective upper limit on the maximum field of view in aperture synthesis.

\begin{figure}[h]
\centering
\scalebox{0.6}{\includegraphics{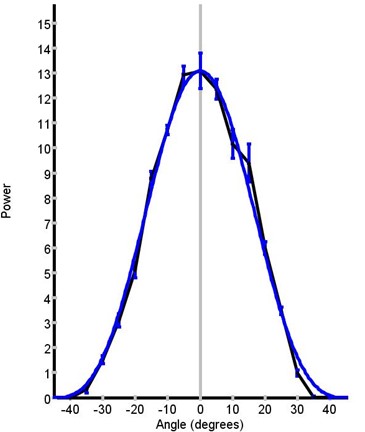}}
\caption{\label{fig:beamplot} A plot obtained by undergraduate students of the primary beam of a VSRT feed. The plot, produced with the VSRTI\_Plotter, shows the detected power as a function of the angular position of a CFL. The angular position is given in units of degrees and the power in units of Kelvins, as is traditionally done in radio astronomy. The model overlay yields a measure of the effective diameter of the antenna feed of approximately 3.8 cm.}
\end{figure}

In the first exercise, students measure the primary beam pattern of the VSRT feeds by placing the active feeds one above the other, making a baseline with zero horizontal length and placing a CFL two meters away at the mid-plane position.  Keeping the feeds fixed, they record data with the CFL placed at various horizontal angles to the mid-plane. The data files can then be drag-and-dropped into the VSRTI\_Plotter ``Plot Beam'' program, producing a graph of the data. The VSRTI\_Plotter program can also overlay a theoretical beam plot, yielding a fitted value of the effective diameter of a feed (see Figure~\ref{fig:beamplot}).

\subsection {A single resolved source}
\label{sec:resolvedsource}

The distance between the antennas (i.e., the baseline) determines the angular sizes of structure in the radio source that the interferometer is most sensitive to.  In short, the baseline acts like a spatial filter.  With longer baselines, the interferometer response becomes more dependent on finer scale structure in the source, while its sensitivity to extended structure decreases. Flux distributed over larger angles is said to be ``resolved out.''  If the baseline is too long, it will not detect the source at all. The detected power of a resolved source, therefore, decreases with increasing baseline length.  One can use the rate of fall-off of detected power to determine the angular size of the source.

In this lab, students use a single CFL located at the mid-plane position between the two feeds and then vary the horizontal separation between the feeds. They produce a plot of the measured power versus the baseline length, finding that the power decreases as the baseline length increases.  In this way students become acquainted with the visibility function, in which the independent variable is baseline length divided by wavelength.

\begin{figure}[h]
\centering
\scalebox{0.8}{\includegraphics{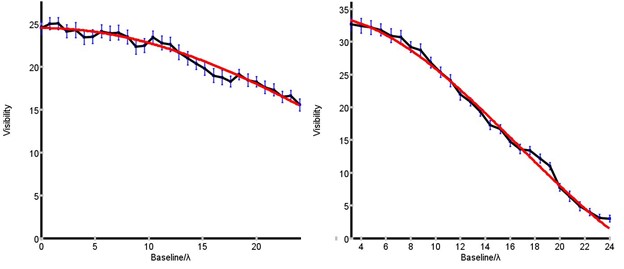}}
\caption{\label{fig:2singlesources}  The detected power from a single CFL as a function of baseline length.  The data show that with increasing baseline length more of the flux of the source is resolved out.
The graph on left results with a smaller source and shows that less of the flux of the smaller source is resolved out than with a larger source. The model curves (shown in red) indicate the angular source sizes were 0.023 and 0.04 radians.}
\end{figure}

Students are instructed to obscure the bottom half of the CFL using a metal plate. They find that the power decreases by about half and verify that this holds at a variety of baseline lengths.  Next, they place two metal plates vertically obscuring the sides of the CFL in order to make a source whose apparent size is smaller. They find that the measured power at short baseline lengths is smaller than for an unobscured CFL, but that the power at long baseline lengths is actually greater.  Figure \ref{fig:2singlesources} displays results of this lab.  The students discover that the rate of decrease of the measured power is inversely related to the angular size of the source.  

\subsection{Double Sources}
\label{sec:pairofsources}

\begin{figure}[h]
\centering
\scalebox{0.8}{\includegraphics{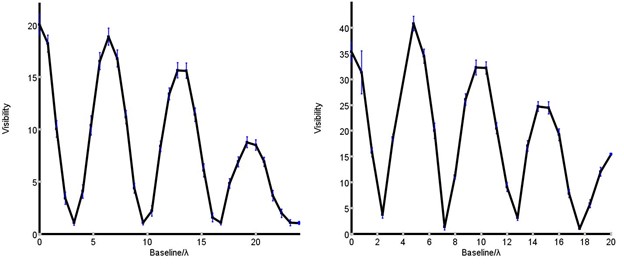}}
\caption{\label{fig:TwoDoubleSources} The visibilities as a function of baseline length for a pair of CFLs.  For the left plot, the CFLs were separated by 30 cm and at a distance of 2 m, while for the right plot the CFLs were separated by 40 cm.  Students discover that the detected signal for a double source has an oscillating dependence on baseline length and that the period of oscillation is inversely dependent on the separation of the sources.  The overall decrease in power with baseline is due to the resolving of each CFL, as seen in Figure \ref{fig:2singlesources}. (The difference in the visibility values between the graphs results because  different lamps were used in the two experiments.)
}
\end{figure}

When the radio source contains two components separated by an angle resolved by the interferometer, the detected power varies with baseline in an oscillatory manner. In this exercise, students discover this by using two CFLs with a fixed separation between them. The students move the VSRT feeds to larger separations and discover that the measured power oscillates with baseline length. The students then move the two CFLs farther apart and repeat their measurements, discovering a reciprocal relationship between the angular separation of the two sources and the distance between minima in the visibility function (Figure~\ref{fig:TwoDoubleSources}).

\subsection{Mystery Source}
\label{sec:mystery}

After completing the exercises discussed in Sections \ref{sec:resolvedsource} and \ref{sec:pairofsources}, the students' new abilities to infer the angular separations and angular sizes from visibilities 
are tested by placing a cardboard box over a pair of CFLs.  After obtaining their data and inferring the source structure, the students can check their answer by lifting the box.

\begin{figure}[h]
\centering
\scalebox{0.5}{\includegraphics{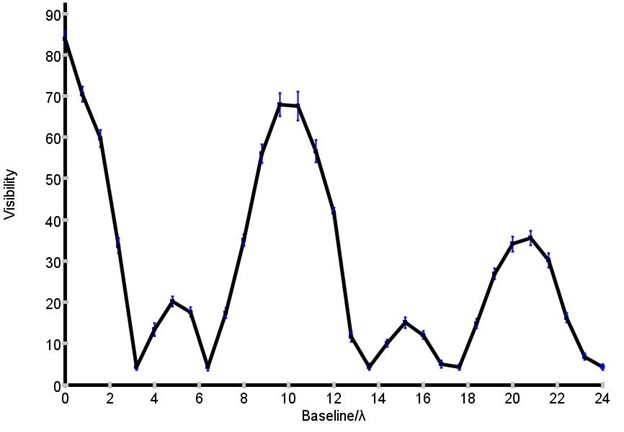}}
\caption{\label{fig:mystery} The detected power as a function of baseline length for a ``mystery source'' hidden inside a box 2 meters away.  The box contained three CFLs with equal separations of 20 cm between the middle and end CFLs, yielding two pairs with angular separations of 0.1 radians and one pair with a separation of 0.2 radians.}  
\end{figure}

A slight increase in complexity of source structure and a challenge to the students' analytical abilities can then be demonstrated with an exercise using three CFLs hidden under a cardboard box.  The instructor sets the CFLs with equal spacings, so that CFLs 1 and 2 and CFLs 2 and 3 create two pairs separated by angle $\theta$ and CFLs 1 and 3 make one pair with a separation of $2\theta$.  The visibility data then show the sum of two sine waves -- one with a larger amplitude and period $=1/\theta$ and the other with a smaller amplitude and half the period (Figure \ref{fig:mystery}).  The students are not informed that there are three CFLs in the box, and are asked to infer, working as team, the source structure considering the principles they learned in the labs. 

\subsection{Fourier Transform}
\label{sec:fourier}

\begin{figure}[h]
\centering
\scalebox{0.4}{\includegraphics{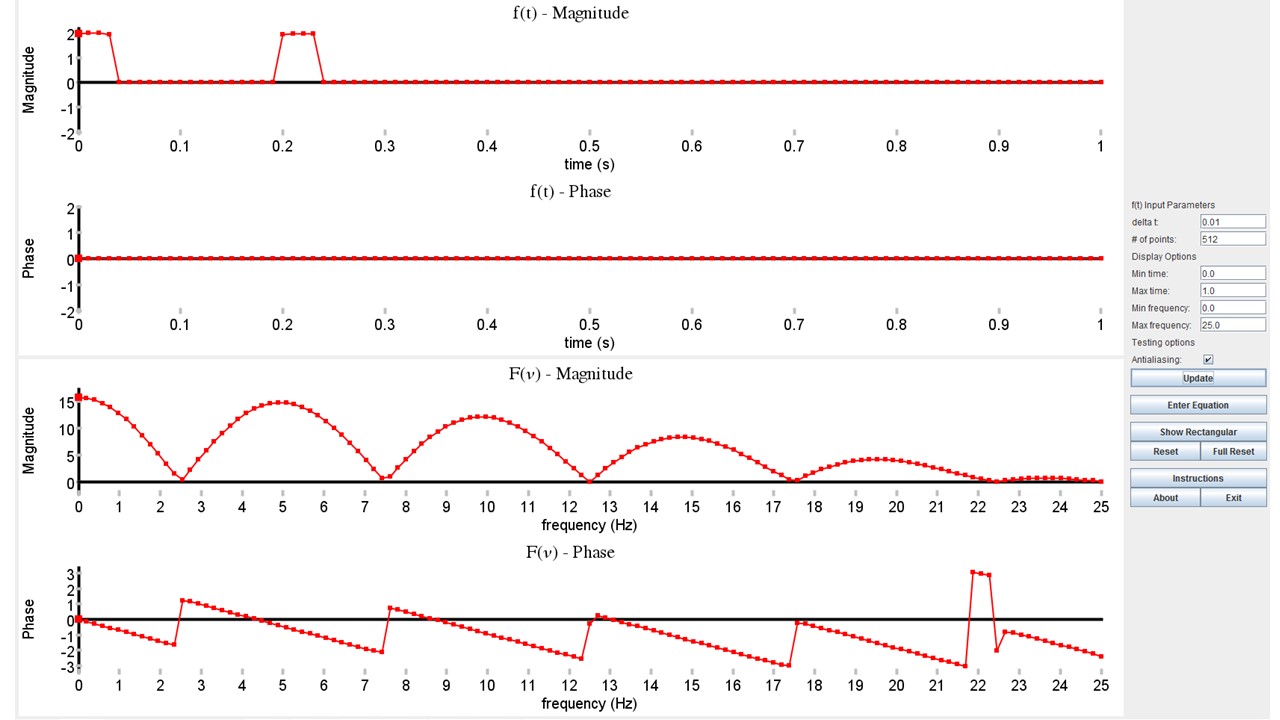}}
\caption{\label{fig:TIFT} TIFT screen shot of the simulation of observation of two resolved sources.  In the ``f(t)-Magnitude'' window, two signals of width 0.03 s and separated by 0.2 are drawn.  The ``F($\nu$)-Magnitude'' window shows that the Fourier transform oscillates with a period equal to 5 and a decreasing envelope and is similar to the visibility function found in the lab observing double sources (see Figure \ref{fig:TwoDoubleSources}).}
\end{figure}

In actual aperture synthesis observations one obtains an image of the source by performing a Fourier transform on the complex visibilities.  Since the VSRT data contain only amplitudes, and no phases, we have provided another java package which students can use to discover the relation between Fourier function pairs.  They find that this relation is identical to that between the visibility function and source structure.  Using the ``Tool for Interactive Fourier Transforms'' (or TIFT), also available online\,\cite{plotter}, students use a simple click and drag operation to make functions representing the brightness distributions of the CFLs in the previous labs.  In a companion plot, they discover that the amplitudes of the Fourier transforms are the same as the visibility amplitudes they measured with the VSRT (Figure \ref{fig:TIFT}).  

\subsection{Solar Diameter}
\label{sec:solar}

\begin{figure}[h]
\centering
\scalebox{0.40}{\includegraphics{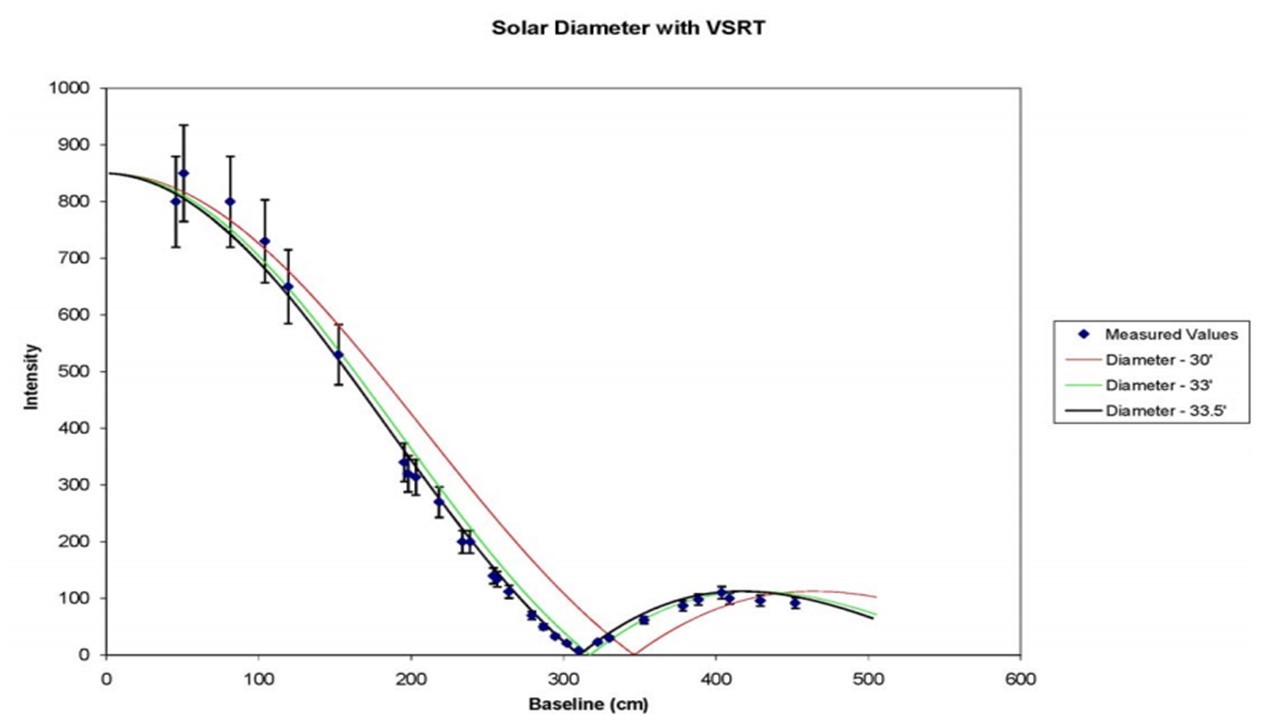}}
\caption{\label{fig:solardia} Plot of the visibilities in a VSRT observation of the Sun, with overlay of curves for 3 different models.}  
\end{figure}

As an exercise using the VSRT to make a measurement of an actual celestial source, the VSRT can also be used to determine the diameter of the Sun.  This involves a slightly more complicated set-up but also provides a practical culmination of the VSRT labs\,\cite{Fish}.  The results of this experiment are shown in Figure \ref{fig:solardia}.

\section{Practical Experience in the Classroom}

Starting in 2012, these labs have been incorporated into the radio astronomy class at Union College.  The class consists of seniors, juniors and sophomores with majors in either physics or engineering.  The coverage of aperture synthesis in the course is accomplished using only the labs; no lectures are provided.  The effectiveness of the labs was tested in the first course they were used.  Both before and after performing the labs, the students were given a quiz on the basics of aperture synthesis, containing the following questions:  

\noindent1.  Explain conceptually how receiving signals with a number of radio telescopes, as with the VLA, contains information about the image of a radio source.

\noindent 2.  What is an ``array'' of telescopes and what are the important criteria in designing the array?

\noindent 3.  What is the ``Visibility function?''

\noindent 4.   Describe what a 1-dimensional visibility function looks like when observing:\\
	a)  a single, unresolved source.\\
	b)  a single, resolved source.  How does the shape of the visibility function change as the angular size of the source increases?\\
	c)   a pair of unresolved sources.  How does the shape of the the visibility function change as the separation of the two sources increases?\\
~	\\
The average score on the quiz before performing the labs was 0.3 points out of 18 (or 1.5\%), demonstrating that none of the students had any prior knowledge about aperture synthesis.  
Afterwards, the average quiz score was 11.9 (or 66\%), indicating a normalized gain (defined by the fraction of the material not known a priori that was learned by the time of the post-test) was 0.66.

The mystery source lab, in which three CFLs were hidden from view, was found to be challenging by the students.  However, with all students collaborating as a team to brainstorm, within a class period and with extensive discussion, the class succeeded in inferring the actual source structure correctly.

\section{Concluding remarks}
\label{sec:conclusion}

These labs have been incorporated in undergraduate radio astronomy courses at Union College and the data for all the plots shown in the figures were obtained by students in these classes.  While single-dish radio observing methods are sometimes included in undergraduate astronomy classes, aperture synthesis is often de-emphasized due to the need to include high-level mathematics.  Using these labs, the students in these classes were able to gain an intuitive understanding that data associated with pairs of antennas at different spacings leads to a function from which one can infer the size of a single source or the separation between sources in the sky.

These labs can be used to stimulate discussions about how the structural details of even a complicated source can be extracted from aperture synthesis data using standard observing setups and imaging algorithms. For instance, after learning that extended sources can be resolved out on long baselines, students can be asked about the importance of matching the angular resolution of an observing array to the scale of the expected structure in an astronomical source and asked to consider why the telescopes in the Jansky Very Large Array and the Atacama Large Millimeter Array are designed to be reconfigurable into arrays of different sizes. After the primary beam lab, students can be asked to speculate why small dishes are used in arrays designed for wide-field imaging, for example.

\begin{acknowledgement}

We are grateful for the significant contributions by Preethi Pratap and Madeleine Needles in
the development of the VSRT, for the careful lab work of the radio astronomy Union College students in the years 2010-2016, as well as George Hassel for ensuring that the labs ran smoothly at Siena. 

The development of the VSRTs was funded by the National Science Foundation CCLI grant DUE-0817136 and the development of the VSRTI\_Plotter Java package by the NSF IIS CPATH award 0722203.
\end{acknowledgement}


\begin{thebibliography}{5}


\bibitem{Nobel}\textit{Nobel Lectures, Physics 1971-1980} (World Scientific Publishing Co., Singapore, 1992).

\bibitem{MSK}Marr, J.\ M., Snell, R.\ L., \& Kurtz, S.\ E., \textit{Fundamentals of Radio Astronomy: Observational Methods} (Taylor \& Francis, Boca Raton, FL, USA 2015).

\bibitem{WRH}Wilson, T.\ L., Rohlfs, K., \& H\"{u}ttemeister, S., \textit{Tools of Radio Astronomy}, 5th ed.\ (Springer-Verlag, Berlin-Heidelberg, 2009), pp.\ 201--234.

\bibitem{BGS}Burke, B.\ F.\ \& Graham-Smith, F., \textit{An Introduction to Radio Astronomy}, 3rd ed.\
(Cambridge University Press, Cambridge, UK, 2010).

\bibitem{TMS}Thompson, A.\ R., Moran, J.\ M., \& Swenson, Jr., G.\ W., \textit{Interferometry
and Synthesis in Radio Astronomy}, 3rd ed.\ (Springer Nature, Cham, Switzerland, 2017), available online at 
https://link.springer.com/book/10.1007\%2F978-3-319-44431-4

\bibitem{Kraus}Kraus, J.\ D., \textit{Radio Astronomy}, 2nd ed., (Cygnus-Quasar, Powell, Ohio, 1986).

\bibitem{AR}Arndt, M.\ B., \& Rogers, A.\ E.\ E.\, \textit{VSRT Introduction} (http://www.haystack.mit.edu/edu/pcr/vsrt-ret/VSRT\_Introduction\_Summer2009.pdf, 2009).

\bibitem{DFN}Doherty, M., Fish, V.\ L., \& Needles, M., The Physics Teacher \textbf{49}, 546 (2011).

\bibitem{VSRTurl}http://www.haystack.mit.edu/edu/undergrad/VSRT/index.html

\bibitem{Fish}Fish, V.\ L., \textit{Measuring the Solar Diameter with the Very Small Radio Telescope} to be submitted to Astronomy Education Review.

\bibitem{MS}Mumford, W.\ W.\ \& Schafersman, IRE Transactions on Microwave Theory and Techniques, \textbf{3}, 12--17 (1955).

\bibitem{plotter}http://www1.union.edu/marrj/radioastro/labfiles.html.
%
%
%
%
%
%
%
%
%

\end{thebibliography}
\end{document}